\documentclass[journal]{IEEEtran}

\usepackage{amsmath,amsfonts,bm}

\def\eqref#1{equation~(\ref{#1})}

\def\1{\bm{1}}

\DeclareMathAlphabet{\mathsfit}{\encodingdefault}{\sfdefault}{m}{sl}
\SetMathAlphabet{\mathsfit}{bold}{\encodingdefault}{\sfdefault}{bx}{n}

\usepackage{cite}
\usepackage{amsmath,amsfonts}
\usepackage{graphicx}
\usepackage{booktabs}
\usepackage{url}
\usepackage{hyperref}
\usepackage{nicefrac}
\usepackage{microtype}
\usepackage{multirow}
\usepackage[dvipsnames]{xcolor}
\usepackage{color,colortbl,arydshln}
\usepackage{enumitem}
\usepackage{circledsteps}
\usepackage{textcomp}
\usepackage[normalem]{ulem}
\usepackage[most]{tcolorbox}
\usepackage{mdframed}
\usepackage{subcaption}
\usepackage{pifont}

\graphicspath{{./images/}}

\definecolor{darkgreen}{rgb}{0 0.7 0}

\useunder{\uline}{\ul}{}

\hypersetup{
    colorlinks=true,
    linkcolor=blue,
    filecolor=magenta,      
    urlcolor=cyan
}

\begin{document}

\title{The Equalizer: Introducing Shape-Gain Decomposition in Neural Audio Codecs}

\author{
Samir~Sadok,
Laurent~Girin,
and Xavier~Alameda-Pineda%
\thanks{Samir Sadok and Xavier Alameda-Pineda are with Inria at Univ.~Grenoble Alpes, CNRS, LJK, France. Laurent Girin is with Univ.~Grenoble Alpes, Grenoble-INP, GIPSA-lab, France.}
\thanks{Samir Sadok and Laurent Girin are corresponding authors and contributed equally. E-mails: samir.sadok@inria.fr, laurent.girin@grenoble-inp.fr.}
}

\maketitle

\begin{abstract}
Neural audio codecs (NACs) typically encode the short-term energy (gain) and normalized structure (shape) of speech/audio signals jointly within the same latent space. As a result, they are poorly robust to a global variation of the input signal level in the sense that such a variation has a strong influence on the embedding vectors at the output of the encoder and their quantization. This methodology is inherently inefficient, leading to codebook redundancy and suboptimal bitrate-distortion performance. To address these limitations, we propose to introduce shape-gain decomposition, widely used in classical speech and audio coding, into the NAC framework. The principle of the proposed Equalizer methodology, easily applicable to any NAC, is to decompose the input signal---before the NAC encoder---into gain and normalized shape vector on a short-term basis. The shape vector is processed by the NAC, while the gain is quantized with scalar quantization and transmitted separately. The output (decoded) signal is reconstructed from the normalized output of the NAC and the quantized gain. Our experiments conducted on speech signals with four different prominent codecs show that this general methodology enables a substantial gain in bitrate-distortion performance, as well as a massive reduction in quantizer complexity. The code is available at \url{https://github.com/samsad35/The-Equalizer}.
\end{abstract}



\section{Introduction}
\label{sec:introduction}

 Speech/audio codecs aim to convert time-domain speech and/or audio signals into compressed bitstreams, and back, for both efficient transmission and storage. After decades of development based on classical signal processing techniques \cite{jayant1985digital, gersho2012vector, kleijn1995speech, spanias2007audio}, neural audio codecs (NACs) based on deep neural networks (DNNs) and trained on massive datasets have emerged as the new de facto standards due to their promising performance. Among the many NACs recently proposed in the literature, we can mention SoundStream \cite{zeghidour2021soundstream}, EnCodec \cite{defossez2022high}, DAC  \cite{kumar2023high}, SpeechTokenizer \cite{zhang2023speechtokenizer}, TS3-Codec \cite{wu2024ts3}, Mimi \cite{defossez2024moshi}, BigCodec \cite{xin2024bigcodec}, WavTokenizer \cite{ji2025wavtokenizer}, and diffusion-based codecs \cite{foti2025design}. Moreover, NACs are increasingly being used as \textit{tokenizers}, providing discrete signal representations suitable for speech and audio generation, spoken language modeling, and many other downstream applications; see recent reviews in \cite{mousavi2025discreteaudiotokenssurvey, guo2025recent}.

A NAC is typically composed of a neural encoder that transforms the input signal into a sequence of latent continuous vectors called \textit{embeddings}, a vector quantization (VQ) module that maps each embedding vector into its quantized version taken in a vector codebook, and a neural decoder that transforms the quantized embedding vectors back to the signal domain. 
This general principle, inherited from the seminal VQ-VAE model \cite{van2017neural}, is illustrated in Block 2 of Fig.~\ref{fig:overall-equalizer}. It is striking to note that most (if not all) modern coders share this general structure, differing essentially in the design of the encoder-decoder and/or the VQ, and possibly in learning strategies.
However, the complex \textit{non-linear} mapping made by the encoder has a direct major consequence that is at the same time trivial, questionable, and, a bit surprisingly, not discussed in the literature (to our knowledge): Two signals that are collinear in the input (waveform) space, i.e.~identical up to a scalar factor, are generally mapped to embedding vectors that are \textit{not} collinear.
As a result,  the latent representation at the output of the encoder entangles the signal amplitude and structural content into a single vector space. This contrasts sharply with the common assumption that, on a short-term basis, speech/audio signals often have a similar general shape (e.g.~spectral content) and a different level (loudness). 
This assumption is commonly exploited in speech/audio processing techniques (see Section~\ref{sec:related-work}) under the general umbrella of \textit{shape-gain decomposition}: a short-term frame of waveform samples (or a corresponding short-term spectrum) is decomposed into a gain, representing its average energy, and a shape vector, representing its normalized direction, envelope, spectral shape, etc. 
Most importantly for our study, shape-gain decomposition is widely used in classic speech/audio codecs, in particular when using VQ: It is processed before the application of VQ to the normalized shape vector and \textit{scalar} quantization to the gain \cite{gersho2012vector, oehler1993mean, kleijn1995speech, gersho2002optimal, spanias2007audio}. 

This \textit{shape-gain VQ} has several major advantages over direct VQ. First, it generally improves the rate-distortion performance. Indeed, let us consider the typical case of several input vectors with a similar general shape and different energy levels. 
With direct VQ, the vector quantizer must allocate multiple codewords to represent what is essentially the same acoustic structure at different loudness levels. This generally results in a very large codebook. 
In contrast, with shape-gain VQ, these vectors are encoded with the same vector codeword plus a scalar gain prototype value that requires a very limited bitrate. Therefore, the size of the VQ codebook can be significantly reduced. This not only generally leads to better rate-distortion performance, but also to a drastic decrease in the complexity of the (classical) codecs, as the search for the optimal codeword in the codebook is often the most computationally demanding operation in signal compression \cite{gersho2012vector}. Finally, the use of normalized vectors has been shown to help stabilize the design of VQ codebooks \cite{yu2021vector}. 

However, to our knowledge, modern neural codecs, including SoundStream, EnCodec, DAC, and many others, have not considered integrating the shape-gain decomposition principle on the input signal. Instead,  as already mentioned, they let the gain and shape of an input vector be encoded jointly in the corresponding embedding vector, leading to suboptimal codebook design, codebook usage, and rate-distortion performance.
In this paper, we address this problem in two steps. First, in Section~\ref{sec:sensitivity}, we report a series of empirical observations that demonstrate and quantify the sensitivity of three prominent NACs, namely DAC, SpeechTokenizer, and BigCodec, to a global gain variation of the input waveform, illustrating the detrimental entanglement of amplitude and shape in the latent space. Second, in Section~\ref{sec:equalizer}, we introduce The Equalizer, a generic methodological framework that explicitly integrates short-term shape-gain decomposition/recomposition and input signal normalization/denormalization at the two extremities of the neural coding pipeline as a solution to this problem. 
Section~\ref{sec:experiments} presents a series of experiments we conducted with speech signals, applying The Equalizer to a set of four different prominent neural codecs, namely EnCodec, DAC, BigCodec, and WavTokenizer. The results demonstrate that the proposed Equalizer method not only ensures robust performance invariance across a wide range of input gain levels but also leads to substantial gain in performance in terms of bitrate-distortion trade-off and complexity reduction.
By revisiting the classical shape-gain decomposition within a modern learning framework, we aim to contribute to bridge the gap between classical signal processing wisdom and contemporary NAC design, combining robustness to input signal variability, improved interpretability, and rate-distortion-complexity efficiency with the flexibility of learned representations.

\section{Related Work}
\label{sec:related-work}

\textbf{NAC design}. The existing work on NACs has essentially focused on the design of the encoder-decoder and of the quantizer.
For the encoder and decoder, a wide range of neural architectures have been explored, including convolutional networks (CNNs) for local feature extraction \cite{kumar2023high}, recurrent models (RNNs, LSTMs, GRUs) for temporal modeling \cite{defossez2022high, zhang2023speechtokenizer}, pure Transformers for long-range dependencies \cite{wu2024ts3}, and hybrid CNN-Transformer architectures \cite{defossez2024moshi}. 
As for the quantizer, its design is generally inherited from classical methods \cite{gersho2012vector}. This includes: single-stage VQ, as used in, e.g., BigCodec~; residual VQ (RVQ), initially proposed as ``multistage VQ'' in \cite{juang1982multiple} and implemented in, e.g., SoundStream, EnCodec, and SpeechTokenizer~; group VQ (GVQ), initially proposed as ``partitioned VQ'', ``split-VQ'', or ``product code VQ'' \cite{gersho2012vector}, and combined with RVQ in a NAC in \cite{yang2023hifi}~; or finite scalar quantization (FSQ) \cite{mentzer2023finite}.
The resulting quantizers can operate at fixed, scalable, or adaptive  bitrates \cite{agustsson2017soft}. They can be learned separately or jointly (end-to-end) with the encoder and decoder, using reconstruction, adversarial, diffusion, or masked prediction objectives \cite{mousavi2025discreteaudiotokenssurvey}. Additional techniques such as stop-gradient and exponential moving average codebook updates \cite{van2017neural} are commonly employed to improve the codebook design and utilization rate. Our work does not address the design of a new NAC or the improvement of existing NAC components (encoder, decoder, or quantizer). We propose a general methodology that is easily applicable to any existing NAC, in an `external' or `internal' manner (see Section~\ref{sec:principle-overview}), to significantly improve its efficiency.  

\textbf{Shape-gain decomposition in classical speech/audio processing.} As an example of extensive use of shape-gain decomposition in classical speech/audio analysis and coding, we can mention the linear prediction coding (LPC) technique. This technique separates the gain and the normalized spectral envelope encoded by a limited set of prediction coefficients \cite{atal1971speech, markel2013linear}. Another example is cepstral modeling, in which the first cepstrum coefficient encodes the energy and the other coefficients encode the normalized spectral envelope on a logarithmic scale \cite{Rabiner1978, benesty2008springer}. 
Many proprietary speech/audio codecs use the shape-gain decomposition. Examples of standardized codecs that use shape-gain VQ include CELP \cite{schroeder1985code}, AMR \cite{bessette2003adaptive}, and Opus \cite{valin2012definition}, to name a few. Note that shape-gain VQ is compatible with the use of a \textit{linear} transform before quantization (and inverse linear transform after quantization), e.g.~the Modified Discrete Cosine Transform (MDCT) in MPEG-2 AAC codecs \cite{bosi1997iso, noll2002mpeg}, as opposed to the non-linear encoder and decoder of NACs (see Section~\ref{sec:principle-overview}). In summary, shape-gain decomposition is a classic of speech/audio processing, including coding, but to the best of our knowledge, it has never been used in NACs.

\textbf{Taking into account the range of input signal level in an NAC.} Training a NAC without properly considering the possibly large range of input signal level in a practical usecase can be considered as a special case of train/test mismatch, since the overall gain of a specific test signal may be absent from the training set. The vast majority of NAC studies simply do not consider this problem and use popular datasets `as is' (in general, the input gain range is limited and similar for train and test signals). To avoid this problem, some studies have considered data augmentation by varying the input level of signals from the training dataset within a given range. For example, a random gain between $-10$ and $6$~dB is applied to normalized signals during training for the streamable version of EnCodec \cite{defossez2022high}. This range is extended to $-24$ to $+15$~dB during the Mimi training \cite{defossez2024moshi}. The authors of SoundStream apply a random gain in the interval $[0.3, 1.0]$ over normalized signals ``to ensure that the model is robust to a wide range of amplitudes'' \cite{zeghidour2021soundstream}. Such a strategy can indeed make the NAC robust to a more or less large input gain range, but still requires a very large codebook to accurately encode all resulting variations of the embedding vectors. In contrast, the proposed Equalizer method normalizes both train and test signals on a short-term basis, implicitly resolving all the above issues and making the encoding/decoding process (almost) totally invariant to the global gain level of the input signal. 

Normalizing the input waveform and transmitting the gain separately has been considered in the non-streamable version of EnCodec \cite{defossez2022high}, but this process is applied on audio segments of 1\,s, representing more than a hundred input/embedding vectors, hence largely beyond the short-term frame level considered in our study. As a result, the input vectors sent to the encoder are not normalized individually and can still have a largely varying norm. A similar long-term normalization/denormalization strategy has been adopted in \cite{du2024funcodec} on 3.2-s audio segments.

\textbf{Normalization of the embedding vectors in an NAC.} 
The authors of DAC enforce an L2-normalization of the latent representation prior to vector quantization and after projection to a low-dimensional space \cite{kumar2023high}, a strategy inspired by a  previous similar approach in image coding \cite{yu2021vector}. Although this removes the explicit magnitude of the latent vectors, it does not remove amplitude information. Instead, it may actually further encourage the encoder to encode both the spectral shape and the energy of the input signal within the direction of the latent vectors. This can result in an even more entangled representation where gain and shape are no longer separable. In contrast, our approach explicitly factorizes gain and shape \textit{prior to the encoder}, preserving a structured and interpretable representation. In short, because of the non-linear nature of the encoder, signal normalization before and after the encoder is very different in spirit and has almost opposite consequences (early disentanglement of shape and gain vs enforced entanglement).

\textbf{Automatic gain control.} Finally, our work is somehow related to automatic gain control (AGC), which is a feedback mechanism widely used in communication and signal processing systems to automatically adjust the signal level so that it remains roughly constant \cite{Rabiner1978, Lyons2011}. Therefore, AGC can be used as a front-end to a NAC. However, AGC is designed to attenuate global abrupt changes in the signal level and, although it can react rapidly depending on the setting of the attack parameter, it provides smooth gain adjustment over time. This  contrasts with our framewise equalization, which works on a short-term basis (see Section~\ref{sec:equalizer}). As a result, AGC does not normalize frame-wise statistics and cannot play the role of the proposed frame-wise equalization.  

\section{An Experimental analysis of the Sensitivity of Neural Audio Codecs to the Input Gain}
\label{sec:sensitivity}

In this Section, we motivate the introduction of the shape-gain decomposition in the NAC framework by conducting an empirical analysis of the effect of an input signal global level variation over the distribution and quantization of the embedding vectors in a representative NAC. The presented experiments aim at answering the following two fundamental questions: (i) Does input gain variation primarily affect the norm, the direction, or both in the latent space? (ii) What is the impact of input gain variations on the stability of the resulting discrete codes?

\begin{figure}[t!]
    \centering
    \includegraphics[width=0.99\linewidth]{figures/problem.png}
    \caption{Effect of global input gain variation on BigCodec, SpeechTokenizer and DAC latent representations.
    Top: Average embedding norm (normalized to $0$\,dB).
    Middle: Average cosine similarity with the $0$\,dB reference embedding.
    Bottom: Discrete code stability after quantization of the embedding vectors.
    The vertical dashed line marks $0$\,dB, the reference gain.}
    \label{fig:placeholder}
\end{figure}

\begin{figure*}[t!]
    \centering
    \includegraphics[width=1.\linewidth]{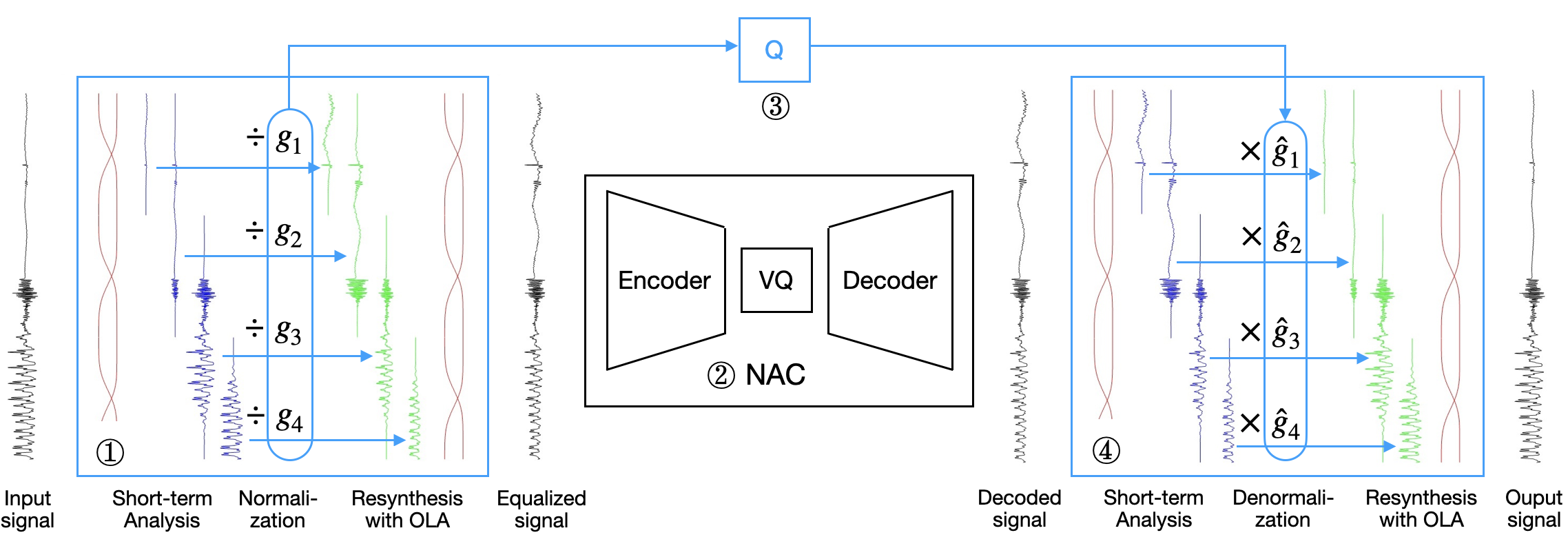}
    \caption{Architecture of the proposed Equalizer NAC pipeline. Block 1: The input signal $\mathbf{s}$ is decomposed into a temporal gain envelope and an equalized waveform $\bar{\mathbf{s}}$ using short-term analysis, normalization, and OLA synthesis. Block 2: The resulting successive shape vectors are processed by a NAC applying vector quantization on the corresponding embedding vectors. Block 3: In parallel, the gain is quantized with scalar quantization (typically $\mu$-law). Block 4: The decoded equalized output waveform and the quantized gain are used to generate the final output waveform using again short-term analysis-synthesis with OLA.}
    \label{fig:overall-equalizer}
\end{figure*}

In these experiments, the input waveform $\mathbf{s}$ of an NAC is modified solely by a global gain factor, while all other signal characteristics are kept fixed, resulting in the gain-scaled version
\begin{align}
    \mathbf{s}_\alpha = 10^{\alpha/20} \cdot \mathbf{s},
    \label{eq:x-gain}
\end{align}
where the gain factor $\alpha$ is expressed in dB and is varied from $-12$ to $+12$\,dB. We observed the behavior of the corresponding embeddings $\mathbf{z} = f_\theta(\mathbf{s})$ and $\mathbf{z}_\alpha = f_\theta(\mathbf{s}_\alpha)$, where $f_\theta(\cdot)$ denotes the encoder.\footnote{In fact, $\mathbf{s}$ and $\mathbf{s}_\alpha$ are segmented in a set of short-term frames that produce a set of corresponding short-term embeddings vectors, but we keep the notation as light as possible in this section.} This was done before and after quantization, and for three different NACs that are representative of the state-of-the-art and for which the source code is available, namely DAC, SpeechTokenizer, and BigCodec. All results presented below are obtained by averaging the metrics over $162$ minutes of speech signals from the LibriSpeech dataset  \cite{panayotov2015librispeech}.

Fig.~\ref{fig:placeholder} (Top) displays the average norm of the embedding vectors $\mathbf{z}_\alpha$ as a function of $\alpha$. For easier visualization and interpretation, the norm of $\mathbf{z}_\alpha$ is normalized by that of $\mathbf{z}$. Fig.~\ref{fig:placeholder} (Middle) displays the average cosine similarity between $\mathbf{z}$ and $\mathbf{z}_\alpha$ as a function of $\alpha$.
First, we can see in Fig.~\ref{fig:placeholder} (Top) that, within the selected interval of gain values, the average embedding vector norm is a non-symmetric monotonic non-linear function of the input gain $\alpha$, indicating that a part of the gain information is explicitly encoded in the magnitude of the latent representations, in a complex manner. Interestingly, the function is increasing for SpeechTokenizer and BigCodec, but it is decreasing, with a lower dynamic, for DAC, indicating that it seems idiosyncratic to the codec. 

Second, and more critically, we can see in Fig.~\ref{fig:placeholder} (Middle) that the cosine similarity between $\mathbf{z}_\alpha$ and $\mathbf{z}$ decreases significantly as the gain departs from the reference level. This is especially true for BigCodec and DAC, with a fast exponential decrease for DAC, and a negative parabola decrease for BigCodec that is slower but reaches lower values at min/max values of $\alpha$ ($0.5$ at $-12$\,dB and $0.4$ at $+12$\,dB). This clearly shows that gain variations in the input signal also induce substantial \emph{directional} changes in the latent space. 
The decrease is more limited for SpeechTokenizer, probably due to the fact that during training its first VQ stage is distilled with Hubert embeddings that mostly carry phonetic information \cite{zhang2023speechtokenizer}, with a joint influence on the encoder due to end-to-end training. This results in embedding vectors that are more independent of the gain compared to the two other codecs. Interestingly, all three curves in Fig.~\ref{fig:placeholder} (Middle) are close to symmetric with respect to the $Y$ axis, which was difficult to guess.  

Finally, Fig.~\ref{fig:placeholder} (Bottom) displays the average discrete code stability (DCS) as a function of $\alpha$. DCS is a measure of the effect of the gain on the quantization of the embedding vectors: it is the percentage of frames for which the codeword indices were different for $\mathbf{z}$ and $\mathbf{z}_\alpha$, i.e.~for which the gain induced a change of codeword. Although they have different dynamics, all three curves look like fastly-decaying exponentials, indicating that even small variations in gain can induce a substantial change in codeword. A DSC of $0.45$-$0.15$ (i.e.~$55\%$-$85\%$ of codeword change) depending on the codec can be observed for only a $\pm4$\,dB gain variation (here also, the curves are surprisingly symmetric). 
This indicates that several codewords are used to encode embedding vectors that correspond to input vectors that differ only by a limited gain variation (note that a variation in either norm, direction, or both of an embedding vector can lead to a change of codeword). In short, we observe a hypersensitivity of the codeword allocation to the input gain, which is clearly not a desired property in VQ.

In summary, our empirical analysis demonstrates (i) that gain and shape are strongly entangled in the latent space of current NACs, and (ii) a strong sensitivity of the NACs (and thus of their codebook design) to the input signal gain. This gain sensitivity is not merely a benign scaling effect, but a structural property that affects codeword allocation and utilization at inference time and codebook design at training time. All this comes in sharp contrast with the shape-gain decomposition principle that underlie classical speech/audio coding, as discussed in the introduction. These findings motivate the need for a global NAC framework that explicitly controls or factors out the input signal gain.

\section{Proposed Method: The Equalizer}
\label{sec:equalizer}

In this section, we present a general methodology, that we call The Equalizer, for exploiting the shape-gain decomposition of speech/audio signals in the NAC framework. We then implement this methodology for four different typical/prominent NACs and show that the proposed method enables a large increase in the performance of these four codecs in terms of bitrate-distortion-complexity trade-off.

\subsection{General Principle and Method Overview}
\label{sec:principle-overview}

As discussed in Section~\ref{sec:introduction} and illustrated in Section \ref{sec:sensitivity}, one notable problem encountered with NACs is that their non-linear encoder entangles the gain and shape information of the input signal into a single embedding vector. If we want to resort to shape-gain decomposition to improve coding efficiency, obviously applying it at the embedding vector quantization level, i.e.~after the encoder, is not an option: This is simply too late. Therefore, we do not consider applying directly shape-gain quantization to the embedding vectors. Instead, we propose applying
the shape-gain decomposition before the encoder, i.e.~at the input signal level. The corresponding shape-gain recomposition is applied at the output of the decoder. The general workflow of the proposed Equalizer methodology is illustrated in Fig.~\ref{fig:overall-equalizer}. 

In some more details, this process is done classically on a short-term frame basis: Each short-term frame of the input waveform is normalized in energy before being sent as an input to the NAC encoder (Block 1 in Fig.~\ref{fig:overall-equalizer}). In this way, input vectors with similar shape and different gains are first transformed to similar normalized vectors before being fed to the encoder, resulting in similar embedding vectors that can be quantized with the same codeword. Note that this requires retraining the NAC on normalized data (see details in Section~\ref{subsec:experimental-setup}). The gain of each short-term frame is quantized separately with scalar quantization and then transmitted to the output of the decoder (Block 3). There, the sequence of quantized gains and the corresponding decoded normalized shape vectors are recombined using overlap-add synthesis to generate the final output waveform with restored energy profile (Block 4).  

Note that in the present study, the shape-gain decomposition and recomposition steps (as well as the gain quantization) are implemented as modules that are external to the NAC. In other words, the shape-gain decomposition and shape vector normalization are not applied directly to the vectors taken as input by the encoder. Instead, a complete waveform is reconstructed from the sequence of normalized short-term frames before being sent to the NAC. We refer to this process as \textit{framewise gain equalization} (hereafter simply \textit{equalization}) of the input waveform.\footnote{Not to be confused with spectral equalization widely used in audio processing.} Similarly, we retrieve the complete decoded equalized waveform at the decoder output and apply \textit{framewise gain restoration}, i.e.~deequalization, to this waveform. This is because of two reasons: First, we do not necessarily have easy access to the input vectors that are actually sent to the encoder and that may result from some specific pre-processing of the input waveform. The same is true for the output vectors just after the decoder. This depends on the NAC, its implementation, and the `transparency' of the code. Second, we want to provide a proof of concept of the proposed methodology that remains general and easily applicable to any NAC with a minimal amount of work. Implementing shape-gain decomposition/recomposition on the actual short-term input/output vectors processed by the encoder/decoder of a specific target codec does not change the foundations of the proposed method. In the case where these vectors are easily accessible, it is straightforward to apply the proposed methodology directly to them. 

Note also that, in this study, we do not implement our methodology in a streaming mode, where all processes in the pipe-line would be applied consistently on a causal frame-by-frame basis, but this is totally feasible without difficulty.

\subsection{Implementation Details}
\label{subsec:implementation-details}

\paragraph{Equalization of the Input Waveform}
Following the standard methodology in speech/audio analysis-synthesis \cite{Rabiner1978, roberts1987digital, spanias2007audio, benesty2008springer}, the input signal $\mathbf{s}$ is first centered and then segmented into overlapping frames $\mathbf{s}_w[m]$ using an analysis window $\mathbf{w}$ of length $N$ and shift $H$: 
\begin{equation}
    \mathbf{s}_w[m] = \mathbf{s}(mH : mH + N - 1) \odot \mathbf{w},
\end{equation}
where $\odot$ denotes element-wise multiplication. 
Then, for each frame, the gain, here denoted $g_m$, is computed as:
\begin{equation}
    g_m = \| \mathbf{s}_w[m] \|_2 = \sqrt{\sum_{n=0}^{N-1} \mathbf{s}_w[m,n]^2},
\end{equation}
where $n$ represents the index of the sample within the frame. This scalar value represents the local energy of the speech/audio signal, and the succession of $g_m$ values over the frames represents its temporal energy envelope (or profile). The normalized shape vector, denoted $\bar{\mathbf{s}}_w[m]$, is then obtained by:
\begin{equation}
    \bar{\mathbf{s}}_w[m] = \frac{\mathbf{s}_w[m]}{g_m + \epsilon},
\end{equation}
where $\epsilon = 10^{-12}$ is a small constant used to prevent numerical instability if $g_m$ is equal to zero.

To reconstruct a complete equalized waveform, which is the signal form intended at the encoder input, the individual normalized signal frames are multiplied by a synthesis window $\mathbf{w}_s$ and recombined using an overlap-add (OLA) procedure \cite{Rabiner1978, roberts1987digital, spanias2007audio, benesty2008springer}:
\begin{equation}
    \bar{\mathbf{s}}(n) = \sum_{m} \bar{\mathbf{s}}_w[m, n - mH] \cdot \mathbf{w}_s(n - mH).
    \label{eq:OLA-norm}
\end{equation}
The settings of the short-term analysis and resynthesis are given and motivated in Section~\ref{subsec:experimental-setup}. The resulting signal $\bar{\mathbf{s}}$ retains the fine time-frequency characteristics of the original input while exhibiting a constant energy profile.
This allows the NAC to focus exclusively on the structural characteristics of the waveform.

\paragraph{Shape Vector Quantization}

The equalized waveform $\bar{\mathbf{s}}$ is sent to the NAC, which will thus process the corresponding sequence of normalized shape vectors and output a decoded equalized signal $\hat{\bar{\mathbf{s}}}$. In this study, we used four different prominent NACs with publicly released code, namely EnCodec, DAC, BigCodec, and WavTokenizer. Our goal being to preserve the versatility of the Equalizer methodology with regard to the codec, the technical details concerning the use of these four specific codecs are given in Section~\ref{subsec:experimental-setup}. Let us simply clarify here that the insertion of an NAC within the Equalizer framework requires its retraining on equalized signals.

\paragraph{Gain Quantization}
In parallel to the shape VQ, the gain $g_m$ is encoded using a non-linear scalar quantizer. Given that human perception of loudness is logarithmic, we apply the classical \emph{$\mu$-law quantization} \cite{jayant1985digital, gersho2012vector, kleijn1995speech, spanias2007audio} to the gain values. The $\mu$-law transformation is defined as:
\begin{equation}
    F(g_m) = \texttt{sgn}(g_m)~\frac{\ln(1 + \mu |g_m|)}{\ln(1 + \mu)},
\end{equation}
where we set $\mu = 255$. This mapping provides a finer quantization resolution for small amplitudes, which is critical to maintaining audio quality in low-energy signal portions. On the decoder side, the inverse $\mu$-law transformation is applied to restore the gain to its original dynamic range.

\paragraph{Output Signal Deequalization}

The final stage of the proposed Equalizer methodology is the \emph{deequalization} of the decoder output signal $\hat{\bar{\mathbf{s}}}$, i.e.~recovery of the original signal level dynamics. This process is similar to equalization of the input waveform, except that here the quantized gain $\hat{g}_m$ is applied to each short-term frame of $\hat{\bar{\mathbf{s}}}$ to restore its original energy (up to quantization errors):
\begin{equation}
    \hat{\mathbf{s}}_w[m] = \hat{g}_m \cdot \left(\hat{\bar{\mathbf{s}}}(mH : mH + N - 1) \odot \mathbf{w}\right).
\end{equation}
Finally, the output waveform $\hat{\mathbf{s}}$ is obtained again using OLA synthesis:
\begin{equation}
    \hat{\mathbf{s}}(n) = \sum_{m} \hat{\mathbf{s}}_w[m, n - mH] \cdot \mathbf{w}_s\left(n - mH\right).
\end{equation}
With proper setting of the analysis-resynthesis parameters, the overall process ensures that the signal maintains high perceptual quality. By decoupling the gain and the shape during all the processing phases and recombining them only at the final step, the Equalizer effectively preserves both the global dynamics and the fine-grained spectral details of the original signal.

\section{Experiments}
\label{sec:experiments}

\subsection{Experimental Setup}
\label{subsec:experimental-setup}

\paragraph{Data}
All experiments were carried out on 16-kHz speech signals. All models were trained using the \emph{LibriSpeech-100} dataset \cite{panayotov2015librispeech}, which contains $100$\,h of speech signals from $251$ speakers. The dataset was randomly divided into $90$\% for training and $10$\% for validation. The evaluation relies on the standard \emph{test-clean} partition ($2,620$ utterances from $40$ distinct speakers, $\approx5.4$\,h) to ensure zero overlap with the training data.  To demonstrate the interest of the proposed Equalizer methodology, we followed the same line as in Section~\ref{sec:sensitivity}, using test signals modified by a global gain $\alpha$ as in (\ref{eq:x-gain}), for a wide range of values, from $-12$ to $+12$\,dB (with a $2$-dB step). Note that the test data at $0$\,dB have the same average level as the training data. In other words, the input gain variation can be seen as a relative deviation from the average level of the training data.   

\paragraph{NACs Choice} In order to show the versatility of the proposed Equalizer methodology with respect to the codec, we conducted experiments with four NACs that are (i) representative of the state-of-the-art, (ii) diverse in terms of bitrate range, coding quality, and architecture, in particular regarding the VQ structure/complexity, and (iii) provided by their authors with an available training recipe so that we can retrain them on equalized signals. EnCodec and DAC are `medium-bitrate' codecs ($4$--$8$\,kbps) that use RVQ, i.e.~multiple cascaded VQ layers, where each successive layer quantizes the residual error of the preceding stage using a different codebook of reasonable size \cite{gersho2012vector, oehler1993mean}. This hierarchical approach enables to drastically reduce the complexity of the quantization at each stage, while ensuring high-fidelity reconstruction when a sufficient number of layers is used. In contrast, BigCodec and WavTokenizer are ultra-low bitrate codecs ($0.5$--$1$\,kbps) based on the projection of the embedding vectors on a low-dimensional space followed by single-stage VQ with a very large codebook.  
We used the implementations provided in the official repositories.\footnote{EnCodec: \url{https://github.com/facebookresearch/encodec}; DAC: \url{https://github.com/descriptinc/descript-audio-codec}; BigCodec: \url{https://github.com/Aria-K-Alethia/BigCodec}; Wavetokenizer: \url{https://github.com/jishengpeng/WavTokenizer}.} We did not make any major modification to the existing code,\footnote{with one exception for EnCodec: Following the architecture proposed in SpeechTokenizer, we have replaced the original two-layer LSTM following the convolution blocks with a two-layer Bidirectional LSTM (BiLSTM), in order to augment the modeling ability of the network by capturing dependencies from both past and future contexts.} we only varied the parameters controlling the bitrate, as detailed below.

\paragraph{Baselines} The objective of this study is not to provide a new, sophisticated, competing NAC with performance at the level of or beyond state-of-the-art performance. Rather, it is to propose a general methodology that is easily applicable to any NAC and demonstrate the potential of this methodology to increase coding performance in terms of bitrate-distortion trade-off and/or complexity reduction. Therefore, for each of the four considered NACs, we quantify the impact of the proposed Equalizer methodology by conducting parallel experiments with a reference baseline that uses the exact same NAC architecture, the same training dataset and training procedure, and the same bitrate configuration as in the corresponding Equalizer configuration, but without shape-gain decomposition. In this baseline, the NAC directly processes the raw input waveform (i.e.~without equalization), and the latent representation jointly encodes shape and gain, as in conventional use. In the following, we call Namecodec-Eq the version of the codec that is combined with The Equalizer, and Namecodec-Base the corresponding baseline.

\begin{table}[t]
\centering
\caption{Configuration of the NACs as used in our experiments.}
\label{tab:codecs}
\begin{tabular}{lcccc}
\toprule
Codec &
Emb.~vec. &
Num.~of RVQ&
Codebook &
Bitrate \\
&
rate ($f$ in Hz) &
stages ($N_q$) &
size ($C$) &
($r$ in kbps) \\
\midrule
EnCodec      & 50 & 8 & 128--1024 & 2.8--4.0 \\
EnCodec*      & 50 & 2, 4, 6 & 1024 & 1.0--3.0 \\
DAC          & 50 & 6, 8, 10, 12 & 1024 & 3.0--6.0 \\
BigCodec     & 80 & 1 & 1024--8192 & 0.8--1.04 \\
WavTokenizer & 40 & 1 & 4096 & 0.48 \\
\bottomrule
\end{tabular}
\end{table}

\paragraph{NACs Configuration} For all codecs used, the bitrate formula is given by:
\begin{align}
    r = f \cdot N_q \cdot \log_2 (C), 
    \label{eq:bitrate}
\end{align}
where $f$ is the embedding vector rate, $N_q$ is the number of VQ stages ($N_q=1$ for BigCodec and WavTokenizer), and $C$ is the size of the codebook(s) (identical for every RVQ stage for EnCodec and DAC). For EnCodec, we set $f=50$\,Hz and varied the bitrate in two manners: In the first one, we fixed $N_q=8$ and varied $C$ within $\{128, 256, 512, 1024\}$. In the second one, we fixed $C=1024$ and varied $N_q$ within $\{2, 4, 6\}$. This second configuration is hereafter denoted EnCodec*. For DAC, we set $f=50$\,Hz, $C=1024$, and varied $N_q$ within $\{6, 8, 10, 12\}$. For BigCodec and WavTokenizer, $N_q$ is inherently fixed to $1$. For BigCodec, we set $f = 80$\,Hz and varied $C$ within $\{1024, 2048, 4096, 8192\}$. WavTokenizer is already limited in coding quality due to its extremely low bitrate, and we simply set $f$ and $C$ to the `default' values $f = 40$\,Hz and $C = 4096$ and did not attempt to decrease $C$. This set-up and the corresponding bitrates are summarized in Table~\ref{tab:codecs}. 

It must be noted that for the baseline codec, the total bitrate is the one given in (\ref{eq:bitrate}), whereas for the Equalizer version there is an additional bitrate due to the coding of the short-term gain $g_m$, equal to $r_g = f \cdot b_g$, where $b_g$ is the resolution of the gain $\mu$-law quantization. In our experiments, we set $b_g = 4$\,bits, which ensures a very limited impact on the coding quality, while leading to an additional bitrate $r_g$ that is assumed to remain relatively small compared to $r$.\footnote{For EnCodec, EnCodec*, and DAC, the additional bitrate is thus $200$\,bps. For WavTokenizer, it is $160$\,bps. For BigCodec, it is not necessary to encode $g_m$ at $80$\,Hz, and we encode it at $40$\,Hz, with a linear interpolation of decoded values for the resynthesis of the output signal. This results in an additional bitrate of $160$\,bps, as for WavTokenizer.} Overall, we will see in the results section that the gain in coding quality and/or codebook size provided by The Equalizer largely compensates for this additional bitrate. 

\paragraph{NACs Training Set-up} To ensure a fair comparison, for each codec, the baseline and the proposed Equalizer versions were trained in the same configuration (the only difference being that, for the Equalizer version, every utterance in the training set was equalized). The implementations for all four codecs were taken from their respective GitHub repositories, and their associated loss configurations were used without any modification. For further details regarding the specific optimization objectives, we refer the reader to the associated original papers. All models were trained across 4 NVIDIA A100 GPUs.

\paragraph{Equalization/Deequalization Set-Up}
For the analysis-resynthesis parameters of the proposed Equalizer methodology, we set $N = 640$ and $H = 320$ for EnCodec and DAC, $N = 400$ and $H = 200$ for BigCodec, and $N = 800$ and $H = 400$ for WavTokenizer. This was to ensure synchronization between the equalized short-term frames of the input signal and the actual vectors taken as input by the NAC encoder (at rate $f$). For all codecs, we used the Kaiser-Bessel Derived (KBD) window with parameter $\beta=4.0$ \cite{princen2003analysis}, for both analysis and synthesis. This choice results from a set of pilot experiments in which we have tested a set of windows that are widely used in speech/audio analysis-synthesis since they ensure perfect reconstruction (PR) of the waveform when we chain analysis and synthesis without modification in between (with a $50\%$-overlap) \cite{Rabiner1978, roberts1987digital, spanias2007audio, benesty2008springer}: KBD, sinus, and rooted Hann. Note that although the PR property is highly desirable in the present coding context, it does not imply that the successive short-term signal frames extracted from the normalized input waveform $\bar{\mathbf{s}}$ sent to the NAC encoder have a norm exactly equal to 1, due to the overlap between the normalized short-term frames in (\ref{eq:OLA-norm}). However, we have observed that the KBD window was systematically providing the lowest standard deviation of this norm (approximately $5\%$ vs $7$--$8\%$ for sinus and rooted Hann). In short, this window is assumed to ensure the best equalization of the encoder input vectors. Consistently, we obtained the best results at the output of the complete Equalizer pipeline using this KBD window.

\paragraph{Evaluation Metrics}
We evaluated the quality of coded speech signals using three widely adopted objective metrics: (ii) Perceptual Evaluation of Speech Quality (PESQ) is a measure of the perceived quality of the resynthesized speech that accounts for factors such as distortion, noise, and other artifacts \cite{rix2001perceptual}; (i) Short-Time Objective Intelligibility (STOI) assesses how intelligible the resynthesized speech is \cite{taal2010short}; And (iii) Mel Cepstral Distortion (MCD) is a widely used objective measure of spectral distortion that computes the average Euclidean distance, in the mel-cepstral domain, between the reference and reconstructed speech signals \cite{Kubichek1993}. Lower MCD values indicate better spectral reconstruction. All three metrics are intrusive, i.e.~they require the original reference speech signal.

\begin{figure*}[t!]
    \centering

    \begin{subfigure}{0.98\linewidth}
        \centering
        \includegraphics[width=\linewidth]{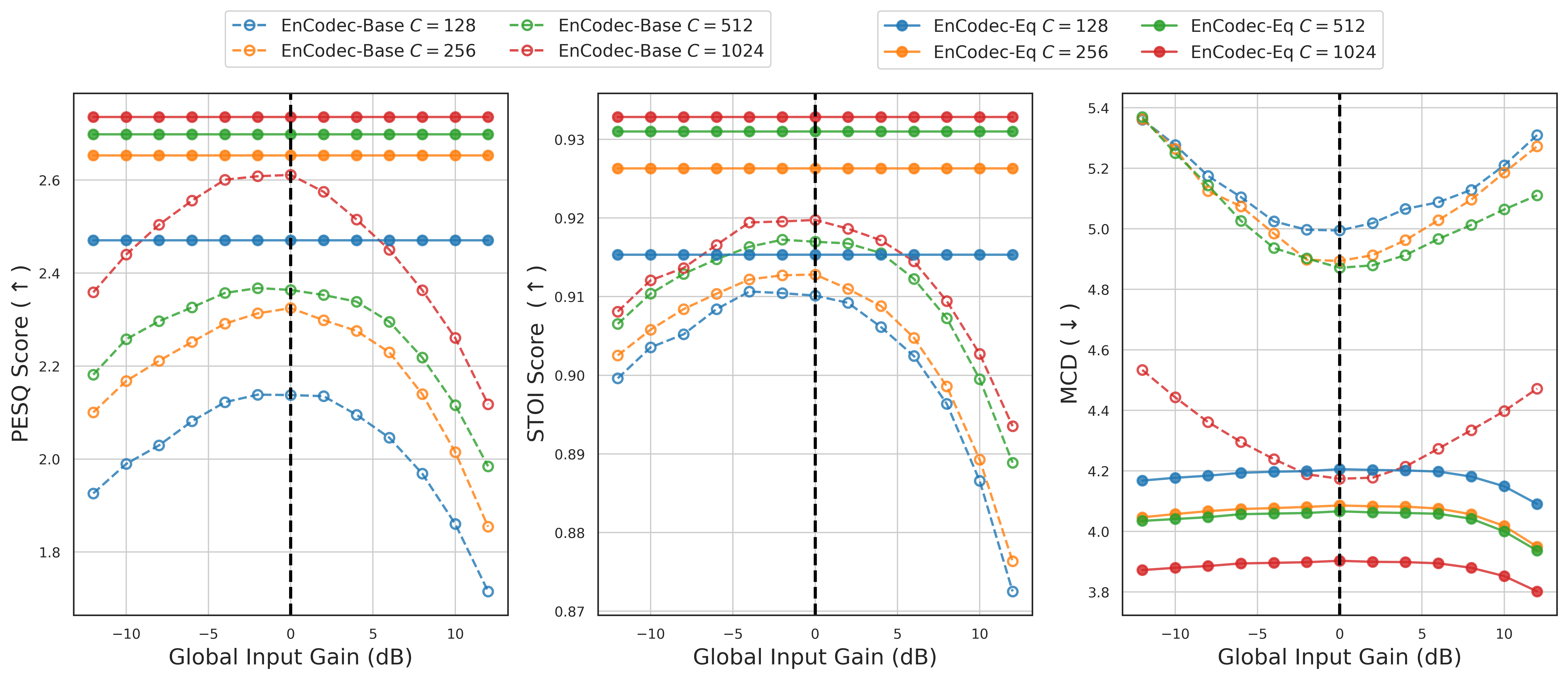}
        \caption{EnCodec ($N_q = 8$).}
        \label{fig:encodec-results-vs-gain}
    \end{subfigure}

    \vspace{0.8em}

    \begin{subfigure}{0.98\linewidth}
        \centering
        \includegraphics[width=\linewidth]{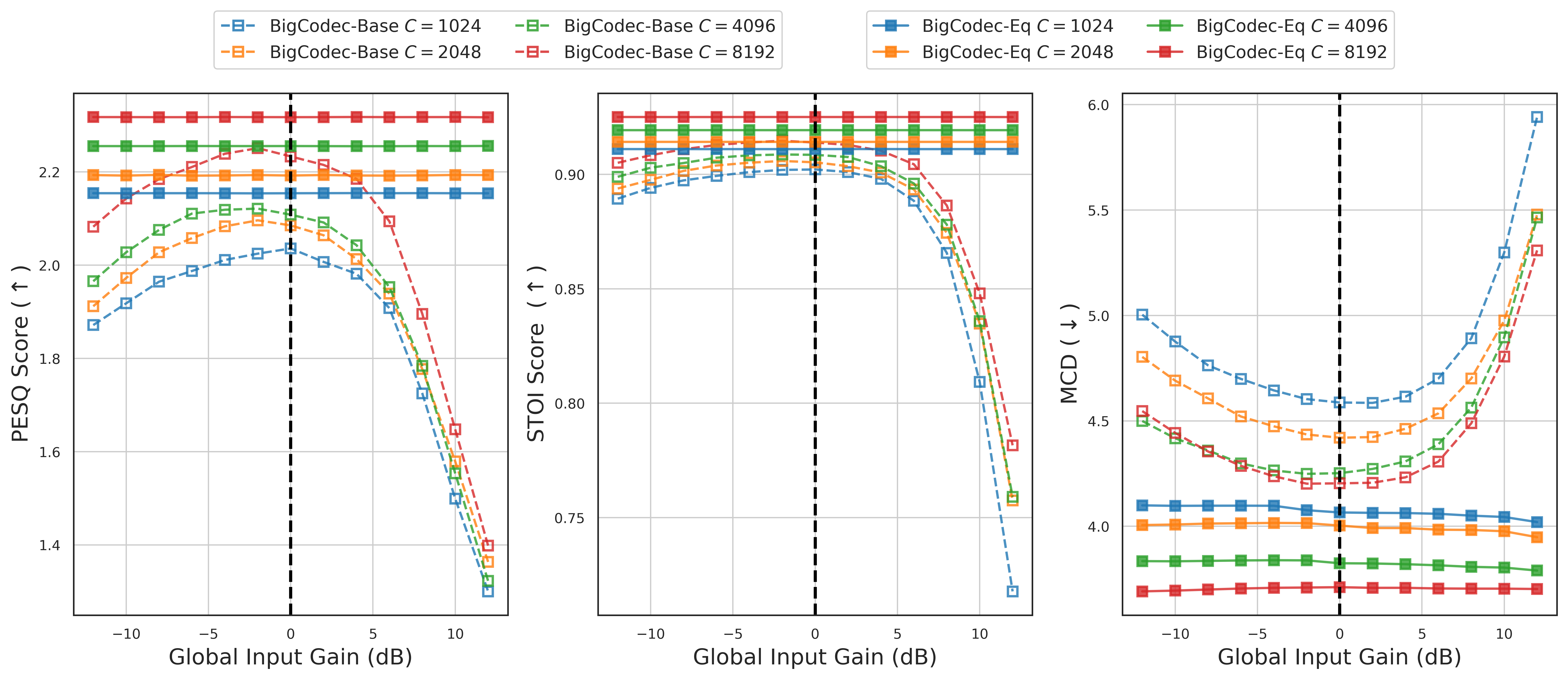}
        \caption{BigCodec ($N_q = 1$).}
        \label{fig:bigcodec-results-vs-gain}
    \end{subfigure}

    \caption{Performance of EnCodec-Eq (a) and BigCodec-Eq (b) (solid lines) and the corresponding baseline codecs (dashed lines) as a function of the input gain $\alpha$, for four different codebook sizes $C$. Scores are averaged over approximately $5.4$\,h of test signals from the Librispeech-test dataset. For both codecs, the Equalizer version remains nearly invariant to changes in $\alpha$, whereas the baseline models degrade as $\alpha$ deviates from $0$\,dB. Furthermore, the Equalizer versions consistently outperform their corresponding baselines even at $0$\,dB.}
    \label{fig:results-vs-gain}
\end{figure*}

\begin{table*}[t]
\setlength{\tabcolsep}{3pt}
\captionof{table}{PESQ, STOI, and MCD scores obtained in our experiments for the four codecs, with and without combination with The Equalizer and for five selected values of the global input gain. All codecs were retrained from scratch on the LibriSpeech-100 dataset. The Equalizer version (\ding{51}) demonstrates near-perfect robustness to input gain variations as opposed to the corresponding baseline (\ding{55}).}
\label{tab:robustness-to-input-gain}
\centering
\resizebox{1.0\linewidth}{!}{ 
\begin{tabular}{cccccccccccccccccc}
\toprule
 & \multicolumn{1}{c}{} & \multicolumn{3}{c}{$-12$\,dB} & \multicolumn{3}{c}{$-4$\,dB} & \multicolumn{3}{c}{$0$\,dB} & \multicolumn{3}{c}{$+4$\,dB} & \multicolumn{3}{c}{$+12$\,dB} \\ 
\cmidrule(lr){3-5} \cmidrule(lr){6-8} \cmidrule(lr){9-11} \cmidrule(lr){12-14} \cmidrule(lr){15-17}
\textbf{Codec} & \multicolumn{1}{c}{\textbf{Equalizer}} & PESQ \footnotesize{$\uparrow$} & STOI \footnotesize{$\uparrow$} & MCD \footnotesize{$\downarrow$}  & PESQ \footnotesize{$\uparrow$} & STOI \footnotesize{$\uparrow$} & MCD \footnotesize{$\downarrow$} & PESQ \footnotesize{$\uparrow$} & STOI \footnotesize{$\uparrow$} & MCD \footnotesize{$\downarrow$} & PESQ \footnotesize{$\uparrow$} & STOI \footnotesize{$\uparrow$} & MCD \footnotesize{$\downarrow$} & PESQ \footnotesize{$\uparrow$} & STOI \footnotesize{$\uparrow$} & MCD \footnotesize{$\downarrow$} \\ \midrule
\multirow{2}{*}{WavTok. (0.48\,kbps)}  & \ding{55} & 1.19 & 0.782 & 7.24 & 1.28 & 0.812 & 6.93 & 1.31 & 0.818 & 6.85 & 1.31 & 0.819 &  6.78 & 1.26 &  0.808 & 6.62 \\ 
& \ding{51} & \textbf{1.58} & \textbf{0.860} & \textbf{5.80}  & \textbf{1.58} & \textbf{0.860} & \textbf{5.85} & \textbf{1.58} & \textbf{0.860} & \textbf{5.87} & \textbf{1.58} & \textbf{0.860} & \textbf{5.86} & \textbf{1.58} &  \textbf{0.860} & \textbf{5.59}  \\ 
\hline
\multirow{2}{*}{BigCodec (1.04\,kbps)}  & \ding{55} & 2.08 & 0.904 & 4.55 & 2.22 & 0.913 & 4.24 & 2.20 & 0.912 & 4.20 & 2.15 & 0.909 & 4.23 & 1.35 & 0.776 &  5.31  \\ 
& \ding{51} & \textbf{2.46} & \textbf{0.930} & \textbf{3.69} & \textbf{2.46} & \textbf{0.930} & \textbf{3.71} & \textbf{2.46} & \textbf{0.930} & \textbf{3.71} & \textbf{2.46} & \textbf{0.930} & \textbf{3.71} & \textbf{2.46} & \textbf{0.930} & \textbf{3.70} \\ 
\hline
\multirow{2}{*}{EnCodec* (3\,kbps)}  & \ding{55} & 2.17 & 0.898 & 4.78 & 2.31 & 0.910 & 4.54 & 2.33 & 0.911 & 4.48 & 2.32 & 0.908 & 4.56 & 1.86 &  0.878 & 4.87  \\ 
& \ding{51}  & \textbf{2.54} & \textbf{0.918} & \textbf{4.14} & \textbf{2.54} & \textbf{0.918} & \textbf{4.16} & \textbf{2.54} & \textbf{0.918} & \textbf{4.17} & \textbf{2.54} & \textbf{0.918} & \textbf{4.16} & \textbf{2.54} &  \textbf{0.918} & \textbf{4.04} \\ 
\hline
\multirow{2}{*}{EnCodec (4\,kbps)}  & \ding{55} & 2.36 & 0.908 & 4.53 & 2.60 & 0.919 & 4.24 & 2.61 & 0.919 & 4.17 & 2.52 & 0.917 & 4.21 & 2.18 &  0.893 & 4.47  \\ 
& \ding{51}  & \textbf{2.74} & \textbf{0.933} & \textbf{3.87} & \textbf{2.74} & \textbf{0.932} & \textbf{3.90} & \textbf{2.74} & \textbf{0.932} & \textbf{3.90} & \textbf{2.74} & \textbf{0.932} & \textbf{3.90} & \textbf{2.74} &  \textbf{0.932} & \textbf{3.80} \\ 
\hline
\multirow{2}{*}{DAC (6\,kbps)}  & \ding{55} & 3.41 & 0.959 & 3.39 & 3.69 & 0.967 & 2.73 & 3.74 & 0.968 & 2.56 & 3.75 & 0.970 & 2.44 & 3.24 &  0.957 & 2.50  \\ 
& \ding{51} & \textbf{3.81} & \textbf{0.974} & \textbf{2.25} & \textbf{3.82} & \textbf{0.974} & \textbf{2.24} & \textbf{3.82} & \textbf{0.974} & \textbf{2.25} & \textbf{3.82} & \textbf{0.974} & \textbf{2.25} & \textbf{3.76} & \textbf{0.973} & \textbf{2.28} \\ 
\bottomrule
\end{tabular}
}
\end{table*}

\subsection{Results}
\label{subsec:results}

\paragraph{Robustness to Input Gain Variations}
Fig.~\ref{fig:results-vs-gain}(a) displays the objective scores obtained by EnCodec-Eq and EnCodec-Base, as a function of the input gain $\alpha$, and averaged over the entire test dataset. For each value of $C$ and each metric, the baseline model (dashed lines) exhibits a clear bell-shaped behavior: a performance peak is obtained near the training condition ($\alpha=0$\,dB) and degrades significantly as $\alpha$ deviates from this value, confirming the sensitivity to input gain variations observed in Section~\ref{sec:sensitivity}. 
For example, for $C = 1024$, applying only a $6$\,dB gain on the input signal leads to a $0.16$ decrease in PESQ value ($2.61$ to $2.45$).
In contrast, EnCodec-Eq (solid lines) shows near-perfect invariance to gain variations, maintaining stable performance over the entire dynamic range.
This shows that shape-gain decomposition effectively removes sensitivity to the global level of the input signal, ensuring consistent reconstruction quality regardless of this level.

Importantly, for each codebook size tested and for each metric, The Equalizer significantly outperforms the baseline at $0$\,dB, that is, even when there is no gain mismatch between train and test data. For example, for $C = 1024$, we obtain PESQ = $2.74$, STOI = $0.933$, and MCD = $3.90$\,dB with EnCodec-Eq vs PESQ = $2.61$, STOI = $0.919$, and MCD = $4.17$\,dB with EnCodec-Base. Other large differences are obtained for $C = 512$, $256$, and $128$. Alternately to quality improvement for a similar complexity, the gain in performance can be interpreted in terms of bitrate and complexity reduction for a similar quality (extended rate-distortion curves will be examined in the next subsection). For example, even at $0$\,dB gain, EnCodec-Eq with $C = 256$ codewords per codebook outperforms EnCodec-Base with $C = 1024$ for all three metrics. This shows that The Equalizer can be used to drastically reduce the complexity of an NAC (here a $4\times$ reduction factor in codeword storage and search; This example also corresponds to a significant bitrate reduction of $600$\,bps, see below). All this confirms the advantage of applying early shape-gain decomposition on speech signals in an NAC framework beyond the induced robustness to gain variation. 

Similar general trends are observed for BigCodec in Fig.~\ref{fig:results-vs-gain}(b), with more asymmetric degradations of the metrics for Bigcodec-Base as $\alpha$ varies. For $\alpha = 0$ dB, BigCodec-Eq outperforms BigCodec-Base with $C = 8192$, achieving higher PESQ and STOI scores, and a lower MCD, while using codebook sizes of only $4096$, $2048$, and even $1024$, respectively.  

We only provide detailed curves for EnCodec and BigCodec, but the effect of the global input gain for all four codecs is reported in a more compact manner in Table~\ref{tab:robustness-to-input-gain} (five values of $\alpha$ and one bitrate configuration for each codec). For EnCodec*, DAC and WavTokenizer, we observe the same general trends as those observed for EnCodec and BigCodec, with a notable improvement of performance for the Equalizer version over the corresponding baseline. The gain in performance is already effective at $0$\,dB and increases as $|\alpha|$ increases, since the baseline performance drops significantly and the Equalizer performance remains remarkably stable (note that the very small variations of a few Equalizer scores across $\alpha$ are due to the $\mu$-law quantization of the short-term gain $g_m$).

\begin{figure*}[t!]
    \centering   \includegraphics[width=1.0\linewidth]{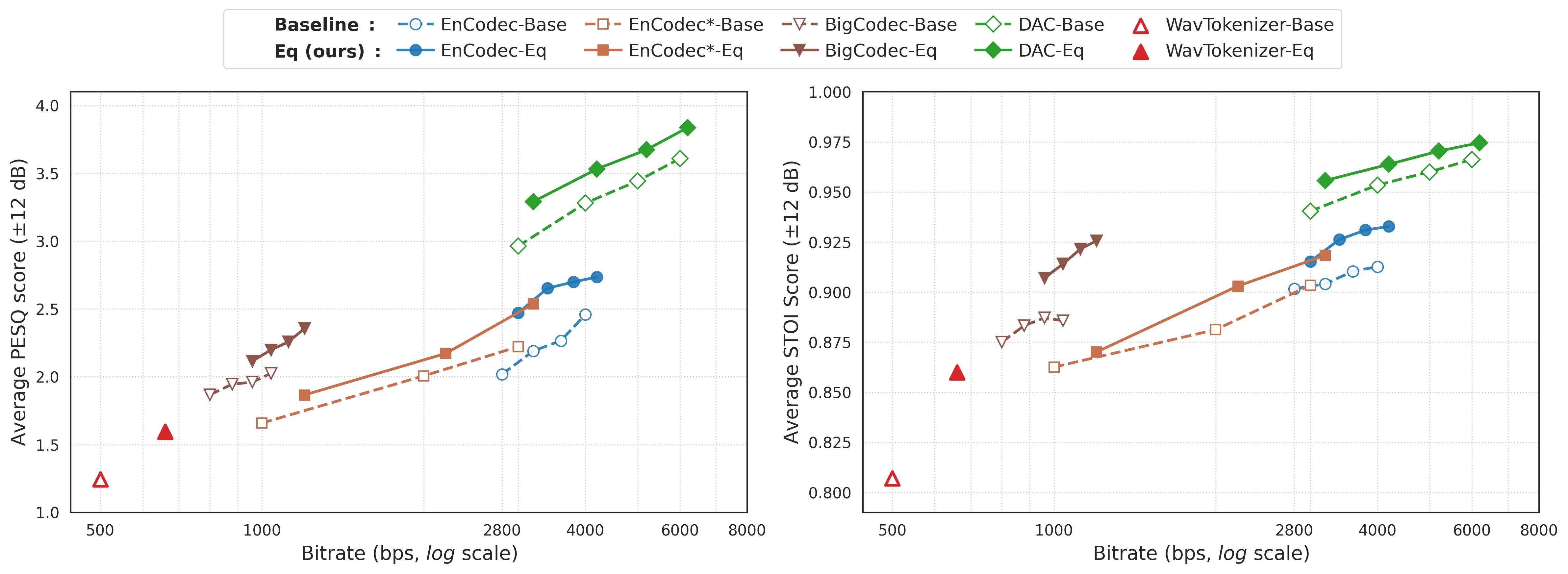}
    \caption{Performance of the proposed Equalizer methodology and of the baselines, as a function of the (total) bitrate, and for the four considered codecs (two versions for EnCodec). Each score is averaged over $\approx 5.4$\,h of test signals from the Librispeech-test dataset and over an input gain range of $\pm12$\,dB. The Equalizer (solid lines) consistently outperforms the corresponding baseline (dashed lines).}
    \label{fig:results-vs-bitrate}
\end{figure*}

\paragraph{Rate-Distortion Performance}
The scores presented above do not completely reflect the performance in terms of bitrate-distortion trade-off, since (i) the variation of $C$ or $N_q$ was not `translated' into bitrate range, and (ii) the proposed Equalizer methodology implies separate coding of the short-term gain $g_m$ and the additional corresponding bitrate $r_g$ (see Section~\ref{subsec:experimental-setup}.d). We therefore now present the results as a function of the total bitrate ($r + r_g$ for The Equalizer vs $r$ only for the baseline). Moreover, we simulate the use of the codecs in a practical scenario where the input gain $\alpha$ can vary in a relatively large range, and we average the metrics across the entire range tested, i.e.~$[-12, 12]$\,dB (and again over the entire test dataset). 

Fig.~\ref{fig:results-vs-bitrate} reports the resulting average PESQ and STOI scores as a function of bitrate. This figure shows that The Equalizer consistently outperforms the baseline for all four codecs and all bitrates, most often by a large margin. In general, for a similar bitrate, The Equalizer achieves substantially higher PESQ and STOI scores compared to the corresponding baseline.\footnote{The measures for The Equalizer are shifted to the right compared to the corresponding measures for the baseline due to the additional bitrate $r_g$, but the Equalizer curves are systematically higher than the corresponding baseline curves.} For example, PESQ = $2.70$ and STOI = $0.931$ for EnCodec-Eq at $3.4$\,kbps vs PESQ = $2.27$ and STOI = $0.910$ for EnCodec-Base at $3.6$\,kbps, hence a $19\%$ improvement in PESQ and a $2.3\%$ improvement in STOI, for a slightly lower bitrate. 
Note that the second configuration of EnCodec (EnCodec*) somehow extends the results obtained with EnCodec to a lower bitrate range. Therefore, the two strategies to vary the bitrate implemented for EnCodec (fix $N_q$ and vary $C$ vs fix $C$ and vary $N_q$) both lead to a significant and consistent improvement of performance when combined with The Equalizer.
For BigCodec, we can compare two configurations with the same bitrate: At $1.04$\,kbps, BigCodec-Base ($C = 8192$) obtains PESQ = $2.03$ and STOI = $0.886$, whereas BigCodec-Eq ($C = 2048$, $r_g = 160$\,bps) obtains PESQ = $2.28$ ($12.3\%$ improvement) and STOI = $0.918$ ($3.6\%$ improvement). For WaveTokenizer, it is more difficult to conclude as the additional bitrate $r_g$ for The Equalizer is no more an order lower than the baseline VQ bitrate. We can still say that for a limited additional bitrate, the gain in PESQ and STOI scores are quite important (resp. $28.1\%$ and $6.6\%$ improvement). At such low bitrate and limited coding quality, it would be interesting to further decrease $r_g$ employing more sophisticated scalar quantization (e.g.~combining $\mu$-law quantization with a predictive coding technique \cite{gersho2012vector}). 

Alternately, for a similar performance, The Equalizer leads to a notable reduction of the bitrate. For example, the PESQ and STOI scores of EnCodec-Eq at $3$\,kbps ($C = 128$) are very close to (actually slightly higher than) those of EnCodec-Base at $4$\,kbps ($C = 1024$). We thus obtain a substantial decrease in total bitrate, $1$\,kbps out of $4$, i.e~$25\%$ relative bitrate gain, for the same coding quality. This confirms that the reduction in bitrate in (each stage of) the RVQ due to the encoding of a normalized shape vector (a total reduction of $1.2$\,kbps = $3$ bits $\times$ $8$ RVQ stages $\times$ $50$ Hz in this example) largely compensates for the additional bitrate ($r_g = 0.2$\,kbps in this example). We can also quickly report another example for DAC: DAC-Eq at $3.2$\,kbps obtains slightly higher scores than  DAC-Base at $4.0$\,kbps, hence $800$\,bps saving ($20\%$).   

In summary, The Equalizer enables either substantial coding quality gain over the baseline for a similar bitrate or substantial bitrate gain for a similar coding quality. All these results confirm that separating the gain and the shape vector allows the vector quantizer to use its capacity more efficiently, as discussed in Section~\ref{sec:introduction}.

\paragraph{VQ Complexity Reduction}
Importantly, the results presented above reflect not only a significant improvement in rate-distortion performance, but also a considerable reduction in VQ complexity. The above example of EnCodec-Eq at $3$\,kbps vs EnCodec-Base at $4$\,kbps corresponds to a $8\times$ reduction factor in codebook size (and thus codeword search), for each of the $N_q = 8$ codebooks. Alternately, the results obtained with EnCodec* show that the number of RVQ stages can be reduced for a similar quality. For BigCodec, in the example given above (with $r = 1.04$\,kbps), we have a $4\times$ reduction factor in the codebook size for the single-stage VQ, together with significantly higher PESQ and STOI scores. Note that this reduction in VQ complexity is already effective even if the Equalizer version keeps the same total bitrate as the baseline version, since the VQ resolution is decreased and the corresponding `saved' bits are `reallocated' to the scalar quantization of $g_m$. It is of course even more effective if the total bitrate of the Equalizer version is further reduced.

\begin{figure}[t!]
    \centering
    \includegraphics[width=1.0\linewidth]{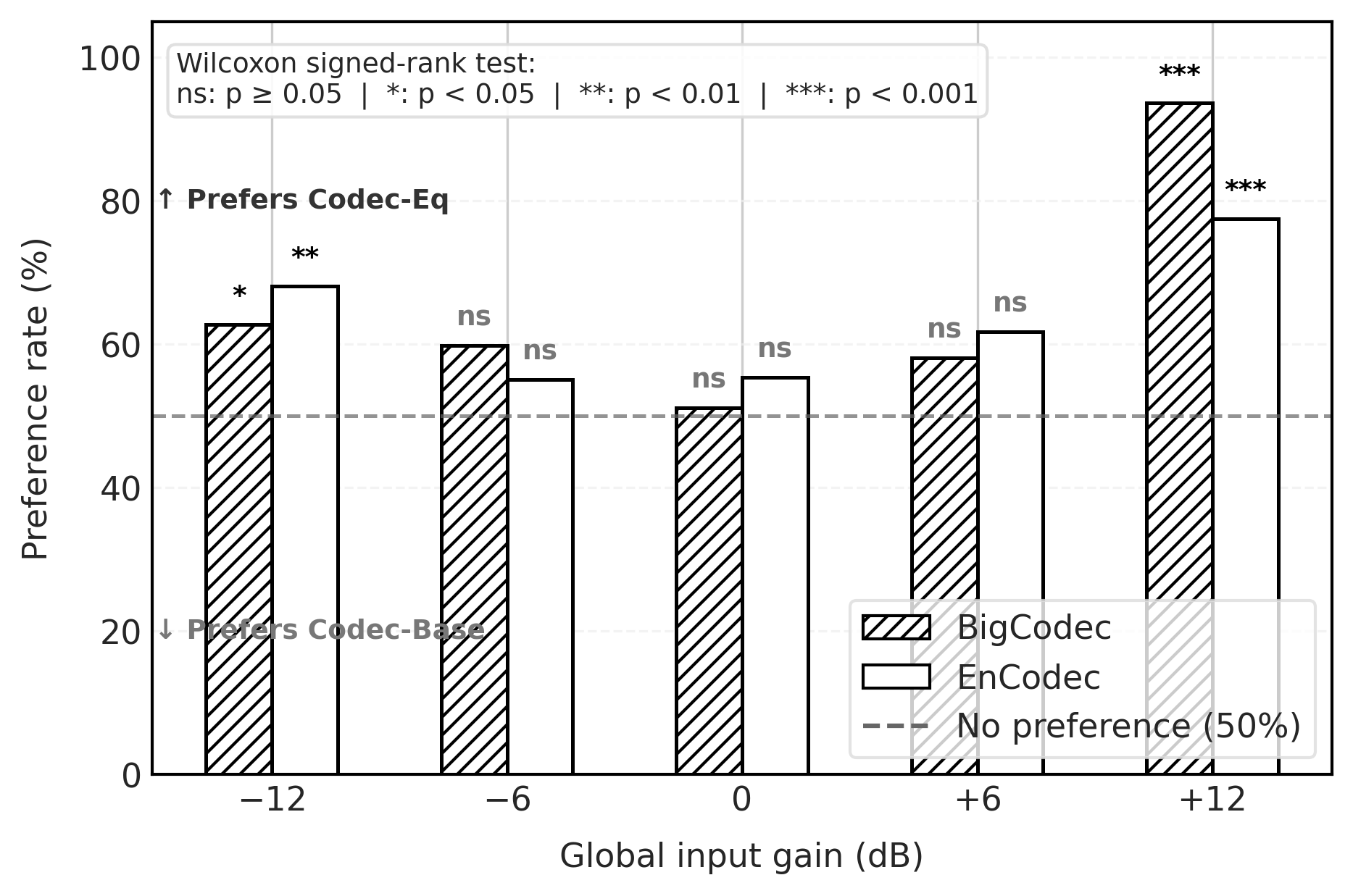}
    \caption{Results of the pairwise preference test.}
    \label{fig:preference-test}
\end{figure}

\paragraph{Perceptive test} To complement the objective evaluation, we conducted a reference-based pairwise preference test, using 4-s speech excerpts randomly selected from the LibriSpeech test-clean dataset. For each excerpt, listeners were presented with three signals: the original reference speech signal, the signal coded by the baseline codec, and the signal coded by the corresponding Equalizer version. The baseline and Equalizer signals were labeled A and B in random order to avoid presentation bias. The participants were instructed to compare the coded signals with the reference and to indicate which of them sounded closer to the reference. They were allowed to replay the three signals as many times as desired before making a decision, although they were encouraged to keep the number of listening repetitions reasonably low. 
Two independent listening experiments were performed, one for EnCodec, with the baseline operating at 4.0 kbps ($C = 1024$) and the Equalizer version operating at 3.4 kbps ($C = 256$, $r_g = 200$\,bps), and one for BigCodec, with the baseline operating at 1.04 kbps ($C = 8192$) and the Equalizer version also operating at 1.04 kbps ($C = 2048$, $r_g = 160$\,bps). 

The results are reported in Fig.~\ref{fig:preference-test}. We used a total of 20 speech excerpts. Each test involved 25 listeners, and each bar in the figure is computed from about $100$ stimuli/answers. For $|\alpha| \neq 0$\,dB and for the two codecs, the Equalizer version was preferred over the corresponding baseline by a majority of listeners. For $\alpha = 0$\,dB, the results are more balanced. Overall, these results are particularly remarkable for EnCodec since, in this test, EnCodec-Eq has a significantly lower bitrate than EnCodec-Base. For BigCodec, the bitrate is identical for both versions, but BigCodec-Eq has a significantly lower quantizer complexity than BigCodec-Base. These perceptual results are globally consistent with the objective evaluation presented above, confirming that the proposed shape-gain decomposition not only improves objective speech quality metrics but also yields a perceptually more faithful reconstruction.

\section{Conclusion}

In this paper, we have presented The Equalizer, a generic methodology for improving the performance of neural speech/audio coders by decomposing the input signal into normalized shape, coded by an NAC, and scalar gain, separately quantized with a scalar quantizer. Although this decomposition has been widely used for decades in classic speech/audio codecs, to the best of our knowledge, it has not yet been proposed within a NAC framework. Importantly, the decomposition is to be performed before the NAC encoder and the corresponding recomposition is to be performed after the decoder, due to the non-linear nature of the neural encoding/decoding processes. In the present study, such equalization and deequalization are implemented as external modules, which can be easily applied to any state-of-the-art NAC, but they can also be applied internally to the actual vectors used as encoder input and retrieved as decoder output. Experiments using four prominent NACs have shown that The Equalizer enables a substantial improvement of the coding performance in terms of bitrate-distortion trade-off together with a massive reduction of the complexity of the quantizer. This  is the mark of a significantly and systemically more efficient VQ codebook design when applying the proposed shape-gain decomposition.

The present work illustrates a broad trend in neural speech and audio coding. While most recent codecs are fully neural, several studies have demonstrated that combining neural networks with carefully chosen signal-processing components can significantly improve coding efficiency. This was proposed for linear prediction \cite{Valin2019LPCNetCodec, zhang2025lspnet}, transform coding \cite{Davidson2024MDCTNet, Jiang2024MDCTCodec, ai2024apcodec}, predictive coding (applied to learned representations) \cite{jiang2023latent, yang2023neural}, or psychoacoustic models \cite{zhen2020psychoacoustic, kim2023progressive}. An overview of these hybrid neural codec approaches is given in \cite{Kim2024NeuralSpeechAudioCoding}. The Equalizer follows the same philosophy, but from a different perspective: instead of modifying the codec architecture itself, it introduces a lightweight, generic signal-processing front-end that can be applied to existing neural codecs. As a consequence, The Equalizer is not only easy to implement, but it is also largely orthogonal and complementary to these hybrid approaches and could potentially be combined with many of them.

The present study focused on equalization and deequalization of the input/output waveforms and its consequences in terms of pure speech coding performances without explicitly examining the effect of equalization on the embedding vectors. The NACs were simply retrained using embedding vectors corresponding to the equalized input waveform, using the same structure and training process as the baseline. Future work will analyze in depth the effect of input vector normalization on the distribution of the embedding vectors and on the geometry of the latent space. We will then explore the use of more sophisticated VQ techniques to exploit the characteristics of this distribution/geometry, with the aim of further improving the overall Equalizer pipeline.

\section*{Acknowledgments} 
This work was granted access to the IDRIS HPC resources under the allocation Grant 2025-A0181016041/2026-AD011016041R2  made by GENCI. It was also funded, in part, by the French National Research Agency (ANR) under project ANR-23-IACL-0006 (MIAI Cluster institute).

\bibliography{mybibfile_improved}

\end{document}